# Implementation and Evaluation of a Cooperative Vehicle-to-Pedestrian Safety Application

Amin Tahmasbi-Sarvestani, Hossein Nourkhiz Mahjoub, Yaser P. Fallah, Ehsan Moradi-Pari, Oubada Abuchaar

*Abstract*— While the development of Vehicle-to-Vehicle (V2V) safety applications based on Dedicated Short-Range Communications (DSRC) has been extensively undergoing standardization for more than a decade, such applications are extremely missing for Vulnerable Road Users (VRUs). Nonexistence of collaborative systems between VRUs and vehicles was the main reason for this lack of attention. Recent developments in Wi-Fi Direct and DSRC-enabled smartphones are changing this perspective. Leveraging the existing V2V platforms, we propose a new framework using a DSRC-enabled smartphone to extend safety benefits to VRUs. The interoperability of applications between vehicles and portable DSRC-enabled devices is achieved through the SAE J2735 Personal Safety Message (PSM). However, considering the fact that VRU movement dynamics, response times, and crash scenarios are fundamentally different from vehicles, a specific framework should be designed for VRU safety applications to study their performance. In this article, we first propose an end-to-end Vehicle-to-Pedestrian (V2P) framework to provide situational awareness and hazard detection based on the most common and injury-prone crash scenarios. The details of our VRU safety module, including target classification and collision detection algorithms, are explained next. Furthermore, we propose and evaluate a mitigating solution for congestion and power consumption issues in such systems. Finally, the whole system is implemented and analyzed for realistic crash scenarios.

## I. Introduction

THE National Highway Traffic Safety Administration (NHTSA) estimated around five million crashes annually and more than thirty thousand fatalities in the United States in 2011. Despite the outstanding efforts of vehicle manufacturers, VRUs such as pedestrians, people with disabilities, cyclists, motorcyclists, public safety personnel, and road workers still account for a significant portion of accident fatalities [1]. Driver distraction and obstruction invisibility, or VRU unawareness of the severity of imminent danger are among the primary reasons behind these fatalities. Although the number of crash victims has been reduced by almost fifty percent since the 1970s, when safety features in vehicles were gradually improved [2, 3], according to the NHTSA, in 2014 traffic crashes caused about five thousand pedestrian fatalities and resulted in more than sixty-five thousand pedestrian injuries in the United States alone [4]. Moreover, another report shows that around seven hundred cyclists lost their lives and about fifty thousand were injured in traffic crashes [5]. In addition, the same report stated that more than one hundred road workers were killed in the United States during 2011.

On the other hand, previous studies show that more than eighty percent of unimpaired traffic accidents are expected to be prevented by safety applications based on V2V and V2I communications [3]. The development of V2V and V2I safety applications and their standardization has been ongoing for more than a decade, and they are anticipated to be deployed within three years [6]. However, such safety applications are severely missing for VRUs.

In this article, we propose, design, and implement an entire V2P framework which complements and improves our previous prototype in [7]. First, existing literature on VRU safety systems is summarized. We afterward discuss the most common and deadliest crash scenarios, to justify our design approach. Subsequently, the proposed V2P framework is described. The VRU safety module is then explained in detail, including target classification and collision detection algorithms. Furthermore, we propose and evaluate a mitigating solution for congestion and power consumption issues in these systems. The results of our outdoor test scenario for the implemented framework is then presented, for which a Qualcomm DSRC-enabled smartphone and a Hyundai-Kia DSRC-equipped vehicle are used (Figure 1). Finally, we provide our conclusions and plans.

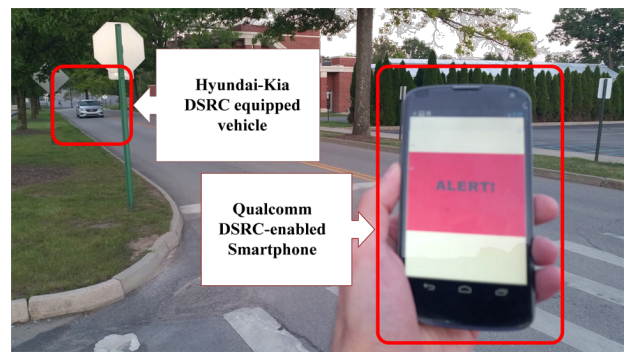

Figure 1- Implemented cooperative VRU safety system

## II. Related Works

VRU safety considerations started around the 1970s by passive safety mechanisms such as head protecting deformable hoods, outside airbags, and helmets, which have become widespread in different parts of current vehicles [8]. Passive safety methods are primarily intended to reduce the severity of the injuries during or after pedestrian crashes, as it is not always possible to avoid accidents. In such incidents, vehicles safety features should safeguard not only their passengers but also VRUs. Therefore, almost all passive safety improvements take place on the vehicle side. Furthermore, VRUs are practically less flexible to be provided with passive safety equipment.

Nowadays, active safety mechanisms, which are capable of predicting and avoiding vehicle to VRU crashes, have become an essential part of advanced driving assistance systems (ADAS) in premium vehicles. Active safety systems consist of sensors to observe the surrounding environment of the vehicle, applications to detect probable hazardous situations, and actuators to respond to them appropriately. A vision-based active safety mechanism for pedestrian detection is described in [9], in which a rangefinder is utilized to detect pedestrians. A similar approach is adopted in [10] which used a camera and laser scanner for pedestrian detection. However, standalone and non-cooperative active safety approaches may only reduce a fraction of accidents due to perception limitation and obstruction of sensors. Most of the sensors have a line-of-sight (LOS) detection capability up to 150 meters and angle of view of about 30 degrees. Thus, occlusion, due to obstruction or severe weather conditions, is their primary constraint. The most important sensor-based non-cooperative VRU safety systems have been categorized and surveyed in [11].

Based on the report published in 2014 by the NHTSA regarding methodologies for Pedestrian Crash Avoidance/Mitigation Systems (PCAM), VRUs who had been obstructed and revealed in less than 2700 milliseconds are challenging and the collision between the VRU and vehicle could not be avoided [12]. Thus, non-cooperative active safety mechanisms are unreliable in scenarios in which VRUs appeared in front of the vehicle in less than 2.7 seconds before impact. Consequently, radio communication technologies are adopted to overcome the occlusion problem, since they can cope with non-line-of-sight (NLoS) situations. Furthermore, vehicular communication technologies typically have an effective range of 300 meters, are omnidirectional, and are capable of two-way information sharing by nature. Hence, communication-based cooperative safety methods provide more accurate and extensive situational awareness in comparison with non-cooperative sensor-based methods [13].

Different strategies have been proposed for cooperative VRU safety applications based on wireless communication technologies. The most significant advantage of cooperative methods is the capability of vehicle and VRU to update their counterpart about their accurate location information periodically. Thus, the performance of cooperative safety application highly depends on the accuracy of global navigation satellite systems (GNSS) such as the global positioning system (GPS), Glonass, or Galileo. A project was performed to evaluate the effect of GPS accuracy on VRUs safety systems. Smartphones are utilized as GPS sensors that provide VRUs' positioning data [14]. According to their results, although sufficient longitudinal accuracy was attained, reliable lane-level localization of VRUs could not be obtained due to high deviations of lateral information.

Tag-based methods which use radio frequency identification (RFID) tags are described in [15, 16]. In these methods VRUs should carry RFID tags and vehicles should mount a transceiver device to detect the presence and predict the movements of the VRUs. However, the effective communication range of RFID tags is limited to 60 meters. Moreover, despite the fact that additional equipment is required to be carried and advanced filtering techniques are needed to detect the VRUs, tags are not capable of providing much information about the pedestrian status.

Undoubtedly, smartphones are the most common portable communication devices which have become an inseparable part of people's everyday life. Furthermore, compared to any other communication equipment, smartphones have full human–machine interfaces (HMIs), powerful processors, and extensive multi-media functionalities. Therefore, cooperative VRU safety based on smartphones is an area of increasing interest. Cellular (UMTS/LTE), Wi-Fi Direct, and DSRC technologies have been the primary candidates to provide the connectivity between vehicles and VRUs in different prototypes.

Cellular technology was used to exchange GPS data amongst smartphone, navigation system of the vehicle, and a central server which is responsible for risk assessment and hazard notification based on provided information [17, 18]. A similar approach has been adopted in [19], in which it was assumed that vehicles and pedestrians had sufficiently precise position information. The same authors extended their work in [20] by comparing different combinations of cellular and ad hoc networks to improve the exchange of information between vehicles, smartphones, and central servers. They also proposed some enhancements to filter unthreatened pedestrians without any justifications. Pedestrians who are not walking toward the road are suggested to be filtered. Other information such as movement history or activity context (e.g., talking, texting, or surfing) is also recommended for more detailed filters. However, their results demonstrated that the communication delay of cellular networks is in the order of seconds which is unbearable for safety critical applications such as cooperative pedestrian safety.

Wi-Fi Direct is a generation of IEEE 802.11 standard which works at 2.4/5 GHz frequency. It enables ad hoc communication of Wi-Fi devices without the necessity of an access point (AP). Wi-Fi Direct has a significant reduction of connection establishment delay, from approximately eight seconds to one second, in comparison with conventional AP-based Wi-Fi. However, in Wi-Fi Direct networks, group owners play the critical role of APs to announce the existence of the network and connect devices to each other. Consequently, as the network formation and group owner negotiation should be renewed if the group owner leaves the network, frequent network reformation might result in an unacceptable delay.

Taking that into consideration, vehicles and VRUs are not acceptable candidates for the group ownership due to their nonstop and fast dynamics. Nevertheless, this technology is not refused by the research community as Wi-Fi-Direct-capable smartphones have been already commercialized. Moreover, Wi-Fi Direct enabled infrastructures could be placed in intersections to resolve the group owner issue, even though less than ten percent of VRU crashes and one percent of fatalities happen at intersections [21]. General Motors has adopted Wi-Fi Direct to prototype a cooperative pedestrian safety system [22]. The same approach to use Wi-Fi Direct is adopted by the authors of [23]. However, their assumption of DSRC and Wi-Fi Direct similarity is not valid. Although they are both part of IEEE 802.11 standard, they are significantly different in many aspects [24]. For example, W-Fi direct suffers from interference from other widely deployed Wi-Fi devices like Wi-Fi hotspot and conventional high-power APs. It also sacrifices the privacy because of IP attached nature of Wi-Fi. Moreover, Wi-Fi Direct is not primarily designed to target high mobility scenarios.

DSRC is a wireless technology specifically designed to support inter-vehicle communications. The main series of standards to support DSRC-based Wireless Access in Vehicular Environment (WAVE) are as follows. The PHY and MAC layers use IEEE 802.11p standard. The IEEE 1609.X family of standards are utilized for the middle of the DSRC stack. Finally, the SAE J2735 standard, which is a dictionary that defines the set of application layer message formats, is placed at the top of this stack. Basic safety message (BSM) is the primary format, which is used in V2V safety applications and contains critical vehicle state information [25]. In the latest version of SAE J2735, the Personal Safety Message (PSM) is defined to convey safety information for different types of VRUs specifically. Furthermore, Optional data elements such as path history can also be included in PSMs if needed [26]. A comparison of BSM and PSM data fields is shown in Table 1.

DSRC is particularly designed for vehicle safety purposes to have low latency and high interoperability. The USDOT is going to mandate the use of DSRC-enabled units in all vehicles, as safety applications based on V2V and V2I communications are greatly capable of preventing traffic crashes [3]. Consequently, DSRC is considered as the most legitimate basis for cooperative VRU safety by standard legislators and technical committees [26]. The main barrier to DSRC-based VRU safety services was equipping the VRU smartphones with DSRC. Recently Qualcomm addressed this concern by announcing their capability to override and upgrade existing Wi-Fi firmware to operate in DSRC band without any additional hardware cost. Cooperative vehicle-to-pedestrian (V2P) safety based on the Qualcomm DSRC-enabled smartphones have been prototyped by Honda and Hyundai-Kia [7, 27]. The effect of pedestrian side DSRC devices on the channel condition for an intersection is studied by the ns-2 simulator [28]. Pedestrian DSRC devices are simulated based on the assumption of being identical to vehicle side devices in all aspect except pedestrian speed, which is not a valid assumption for real world scenarios as smartphone limitations introduce additional constraints to DSRC devices. A similar simulation-based study using ns-3 simulator is also presented in [29]. To the best of our knowledge, our proposed method is the first system design and implementation of DSRC-based cooperative V2P system founded on the SAE safety requirements, specification, and standards.

Table 1- SAE J2735 BSM and PSM data fields comparison

|  | BSM | PSM |
|---|---|---|
| Common Fields | Message Count | Message Count |
|  | Temporary ID | Temporary ID |
|  | DSecond | DSecond |
|  | Position 3D | Lat., Long., Elev. (Position 3D) |
|  | Positional Accuracy | Positional Accuracy |
|  | Speed | Velocity |
|  | Heading | Heading |
|  | Acceleration Set 4-Way | Acceleration Set 4-Way |
| Specific Fields | Transmission State | Personal Device User Type |
|  | Steering Angle | Personal Device Usage State |
|  | Brake Status | Personal Crossing Request |
|  | Vehicle Size | User Size and Behavior |

### III. PEDESTRIAN CRASH SCENARIOS

While vehicle to VRU accidents could happen almost anywhere, the NHTSA reports the following as the most common and injury-prone crash scenarios [21]:
1) Pedestrian crossing the road in front of straight going vehicle (Figure 2-a).
2) Pedestrian crossing the road in front of the right-turning vehicle at the intersection (Figure 2-b).
3) Pedestrian crossing the road in front of the left-turning vehicle at the intersection (Figure 2-c).
4) Pedestrian walking beside the road and vehicle is going straight (Figure 2-d).

Although these four scenarios account for only 46% of all crashes, they are reported for 98% of fatalities, injuries, and monetary damages of all pedestrian crashes. Therefore, the focus of our VRU cooperative safety module stems from these four scenarios. Based on the NHTSA report, 78% of accidents happen at non-intersection scenarios when pedestrians are crossing improperly, running onto the road, or being distracted. On the vehicle side, 87% of accidents occur when vehicles are going straight, and 10% of them are at intersections and vehicles are turning left or right [1].

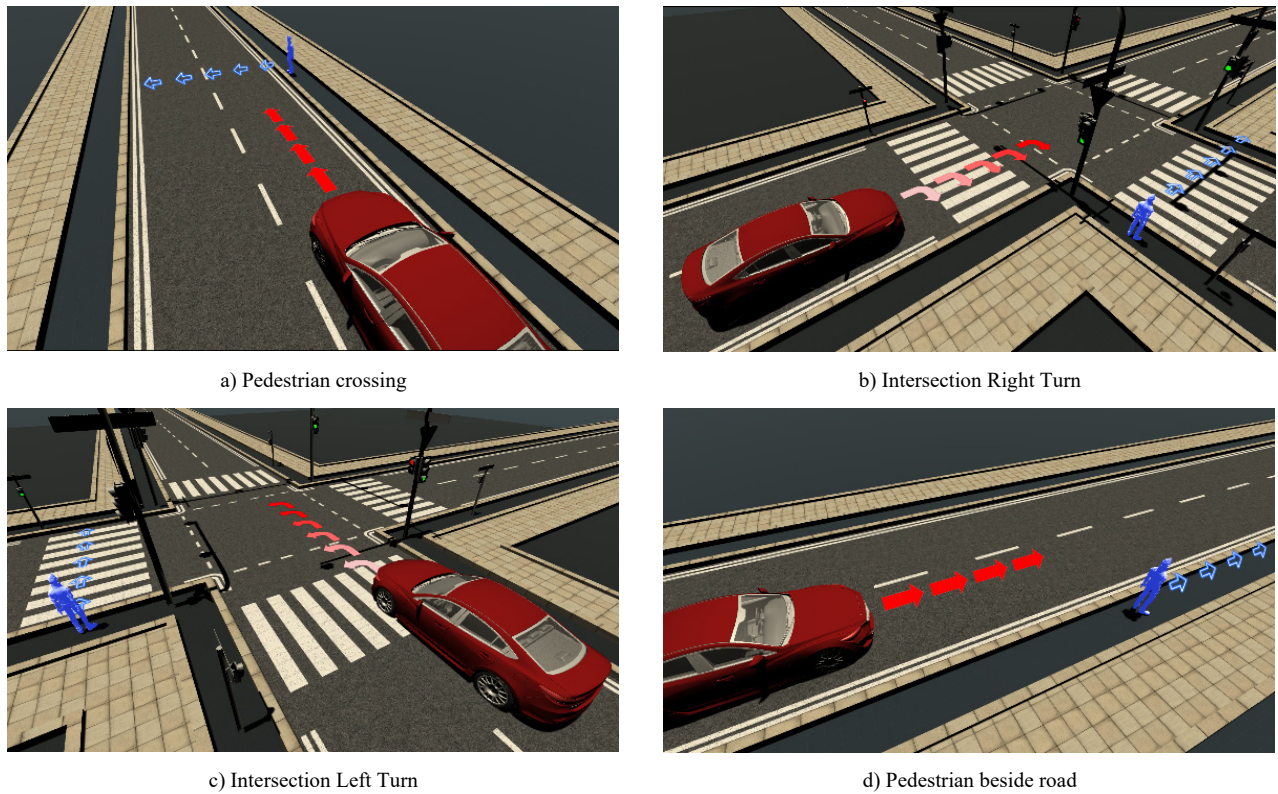

Figure 2- the most common and injury-prone vehicle-pedestrian crash scenarios

The most common crash scenario is the first one in which a crossing pedestrian is hit by a straight going vehicle. The second and third scenarios are common turning accidents at intersections. Although turning accidents consist of the complicated vehicle and pedestrian maneuvers, the injuries are less intense. Finally, the fourth scenario, which is a fatal case, involves fast and unforeseeable maneuvers in which pedestrians are walking along the street without even intersecting the path of vehicles.

## IV. Proposed V2P Framework

As mentioned before, obstructed VRUs cannot be timely detected by sensors such as radars and cameras. Therefore, communication-based cooperative methods with omnidirectional and extended situational awareness range should be employed to address such hazardous scenarios.

Our proposed V2P framework is shown in Figure 3, in which the vehicle and VRU are equipped with a wireless safety unit (WSU) and a DSRC-enabled smartphone, respectively. The WSU is connected to the controller area network bus (CAN bus) and a GPS receiver. The CAN bus continuously provides the WSU with all sensory data collected by local sensors, such as the odometer, accelerometer, brake status, turn signals, and radar, which express the current state of the vehicle. The GPS receiver periodically feeds the location information into the WSU and might also support offline Geographical Information System (GIS) maps, like Google maps, which can help in identifying static structures and road geometries. Moreover, both the WSU and the smartphone contain a DSRC communication module which is capable of communicating SAE J2735 messages such as basic safety messages (BSMs) and personal safety messages (PSMs). Subsequently, all of this information is fed into the situational awareness subsystem to create the extended real-time map of the surrounding environment.

The situational awareness subsystem consists of three components: map-update module, surroundings real-time map, and tracking module. The map-update module is in charge of refreshing the real-time map with the latest received information from sensors, GPS, and DSRC module. Thus, the real-time map database keeps the records of its surrounding entities, such as all neighboring vehicles, VRUs, and other detected objects. Every map record consists of the latest available information about a particular neighboring entity, such as its position information, speed, size, or type of VRU. The map is created and updated based on the latest self-position as the origin and the latest heading direction as the x-axis of its coordinate system. In order to formulate this coordinate system, GPS information, which is in the World Geodetic System 1984 format (WGS 84), is converted into Earth-Centered, Earth-Fixed format (ECEF). The ECEF format is offset to the center of the vehicle on the ground and then transformed into East-North-Up (ENU) coordinates. The ENU coordinates are finally rotated toward the latest heading direction. Additionally, the tracking module utilizes the previous and current map records to construct the path history points and path prediction points for all map elements. This information is then used by safety subsystems to identify pre-crash scenarios. Furthermore, the situational awareness subsystem is responsible for delivering the latest information to the communication module to be broadcast to other vehicles and VRUs.

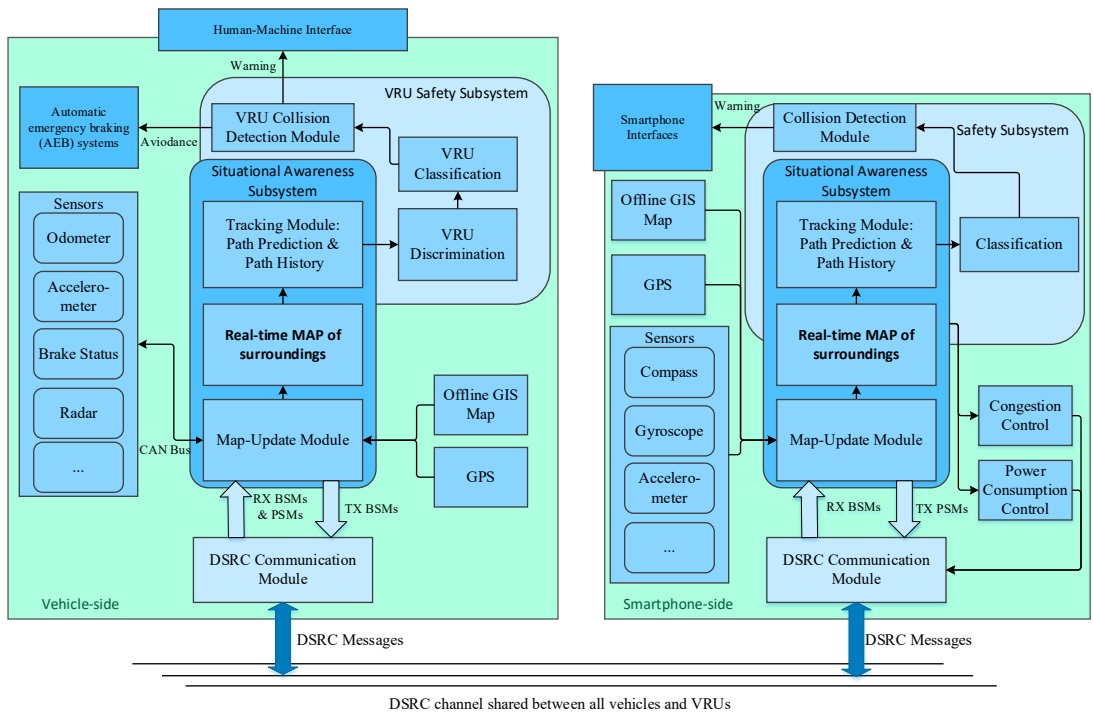

Figure 3- Proposed V2P framework

On the vehicle-side, the VRU safety subsystem utilizes the results of the situational awareness subsystem to determine possible hazardous situations and generate appropriate notifications and proper signals. This subsystem consists of VRU discrimination component, VRU classification component, and VRU collision detection module.

The VRU discrimination component distinguishes between different types of VRUs, as particular VRUs may require different classification or collision detection approaches. This requirement is also considered in PSM that has a field for device user type, which describes the kind of non-vehicular road user whose condition information is being transmitted. These types have been exemplified in SAE J2735 as pedestrians, pedal cyclists, public safety workers, and animals. Nevertheless, this information might be unavailable explicitly, especially when the device is a smartphone since they can be carried by pedestrians, cyclists, road workers, and even vehicle passengers. Therefore, whenever the field is blank, both the smartphone and the counterpart vehicle should detect and identify the VRU type to the greatest extent.

Although numerous PSM fields could be helpful in various situations, the most informative candidates are velocity, four-way acceleration set, path history, and path prediction. Velocity could help to differentiate between pedestrians, pedal cyclists, and motorcyclists. It can also speculate the age, disability, and impairment of pedestrians. Four-way acceleration set, which is the set of lateral, longitudinal, and vertical accelerations along with yaw rate could also be used to discriminate between pedestrians and motorized road users. Path history and path prediction of a VRU could determine whether the VRU is crossing the road or staying within the boundaries of the road, which not only can separate riders from pedestrians but also public safety and road workers from other pedestrians. After the VRU type has been determined, the smartphone can include it in the PSM, and the vehicle VRU discrimination component can feed it into the VRU classification component.

The VRU classification component executes a target classification algorithm to label safe, risk, or danger zones based on current vehicle dynamics (Figure 4). This information is then used by the collision detection module to generate timely warnings or activate collision avoidance actuators. The warnings are then interpreted by the HMI and could be in the form of visual, auditory, or haptic alarms. The collision avoidance signals trigger evasive mechanisms such as automatic emergency braking (AEB) or crash imminent braking (CIB). The details of our proposed algorithm for VRU classification and collision detection module is provided in the next subsection.

The safety subsystem of the smartphone is designed conceptually similar to the VRU safety subsystem of the vehicle, except that it does not require the discrimination component, has different HMI capabilities, and runs a slightly modified detection algorithm. In addition, the smartphone-side contains two extra modules, namely congestion control and power consumption control. The congestion control module mitigates the communication load of the DSRC network in crowded and dense areas, and the power consumption control module reduces battery usage by minimizing the overall safety system duty cycle.

## A. Target Classification and Collision Detection Algorithms

We design the target classification and collision detection algorithms to separate and alert the vehicles or VRUs based on their current movement patterns if they might encounter an accident. The classification, detection, and alerting should be prompt enough to prevent crashes while accounting for the driver reaction delay. However, too early alarms may also be treated as false positive and breed mistrust toward the whole system.

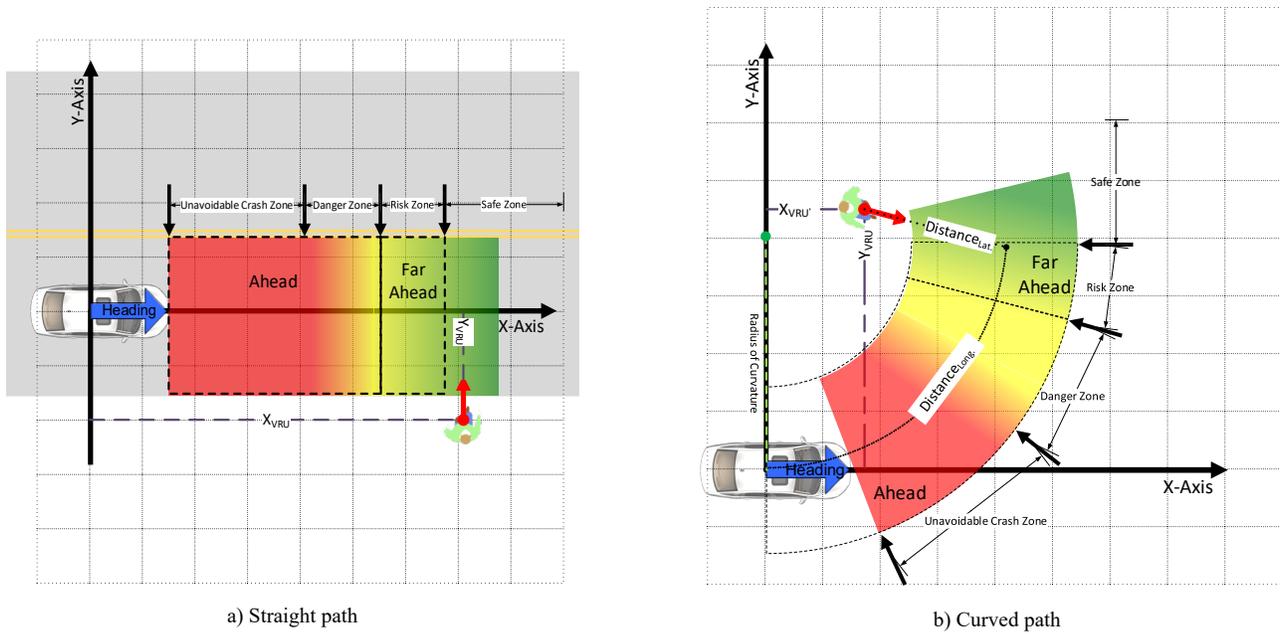

a) Straight path             b) Curved path

Figure 4- Different severities of possible crashes in straight and curved paths

Taking into consideration the kinematic constraints of vehicles and hazardous areas around them, we define different zones based on the severity of possible crashes in our target classification algorithm. Figure 4 illustrates these zones conceptually, in which the width of each zone is equal to lane width, and the length of it at each moment is dependent on the speed and acceleration of the vehicle. The target classification algorithm has the responsibility of assigning the corresponding zone labels and updating them periodically.

The unavoidable crash zone is the area inside which the accident is inevitable, even with the most immediate harshest possible brake. The minimum time-to-stop ($TTS_{min}$) is defined as the time it takes to stop the vehicle with maximum braking deceleration ($d_{max}$), based on current speed ($v_{cur}$), current acceleration ($a_{cur}$), and driver reaction delay ($T_{DRD}$).

$$TTS_{min} = -\frac{v_{brk}}{d_{max}} + T_{DRD} \qquad (1)$$

where $v_{brk}$ is the speed of the vehicle after driver reaction delay and equals to $v_{brk} = a_{cur} \times T_{DRD} + v_{cur}$.

The corresponding traveled distance, which is defined as the minimum distance-to-stop ($DTS_{min}$), determines the length of the unavoidable crash zone and can be calculated as

$$DTS_{min} = D_{DRD} + \left(-\frac{v_{brk}^2}{2 \times d_{max}}\right) \qquad (2)$$

where $D_{DRD}$ is the traveled distance during driver reaction delay

$$D_{DRD} = \frac{1}{2} \times a_{cur} \times T_{DRD}^2 + v_{cur} \times T_{DRD} \qquad (3)$$

Coming after is the danger zone, in which a certain forthcoming accident is avoidable if a timely warning is generated for the driver. Such warning is known as an *imminent warning* and could be followed up by an AEB or CIB signal to prevent the crash. The length of the danger zone is specified by the distance traveled during $T_{iw} = TTS_{min} + T_{guard}$, which is denoted by the guard distance-to-stop ($DTS_{guard}$). $T_{guard}$ determines how soon the first imminent warning should be generated after a crash is detected, assuming that the driver reaction delay is already considered in $TTS_{min}$. Therefore, the level of conservatism of method can be adjusted by $T_{guard}$, considering that there is always a trade-off between the level of conservatism and too early disruptive false alarms.

Despite this trade-off, the driver should still be informed and advised about a probable accident to react moderately with peace of mind, which results in a smoother braking behavior and more comfortable ride. Accordingly, the region for which the driver should only be advised about the possibility of an accident with an *advisory warning* is called the risk zone. Therefore, the length of the risk zone is dependent on the time it takes for the vehicle to stop with a moderate braking deceleration ($d_{mod}$), considering driver reaction delay and current vehicle velocity and acceleration. This duration is denoted by $T_{aw} = T_{iw} + T_{mod}$, in which $T_{mod}$ is added as a configurable parameter for the sake of gentle braking. The corresponding travelled distance is defined as the moderate distance-to-stop ($DTS_{mod}$), which explains the length of risk zone. The regions beyond these zones are considered as safe.

The dimensions of each zone should be recalculated periodically (every 100 milliseconds). Each VRU is afterward classified in either the safe, risk or danger zone based on its estimated relative lateral and longitudinal positions provided by the tracking module, which is the path prediction of the VRU.

ITSM-16-08-0111.R2                                                                                                                                            7Subsequently, advisory or imminent warnings are generated for them by the collision detection algorithm. The lateral and longitudinal distances are denoted by $D_{lat}$ and $D_{lon}$ and computed based on the projection of VRU's predicted position on the path of the vehicle. The distance between the VRU predicted point and projected point is $D_{lat}$ and the distance between the projected point and the vehicle center point is $D_{lon}$. As it is shown in Figure 4-a, since in straight road scenarios the heading of vehicle is in the same direction with its path, the $D_{lat}$ and $D_{lon}$ are equal to $Y_{VRU}$ and $X_{VRU}$, respectively. On the other hand, for curved paths, such as curved roads or turning maneuvers, the heading is tangent to the curve of vehicle path as shown in Figure 4-b. The radius of this curvature ($R$) is derived by $R = \frac{\psi}{v}$, where $\psi$ and $v$ are yaw rate and velocity, respectively. The intercept of predicted path of VRU with the curve of vehicle path is then used to calculate the lateral and longitudinal distances.

Note that, due to the deterministic path prediction and path history content of the BSMs, the proposed classification and detection algorithms are also designed in a deterministic manner. However, probabilistic detection methods can be designed on the same framework to consider localization accuracy and natural human motion patterns. For example, probabilistic solutions for frontal collision detection were introduced in [30-32].

*B. Congestion Control and Power Consumption Control Mechanisms*

Although the effect of congestion and power consumption issues is not visible in proof-of-concept and prototypes of V2P systems, these two subjects along with smartphone positioning inaccuracy are the main challenges in front of realizing a large-scale ubiquitous V2P safety system.

Congestion, which is also investigated in V2V safety systems as the scalability problem, happens when multitudinous communication devices concurrently utilize the wireless channel in the range of each other. As a result, the channel throughput drops drastically, if the rate and range of transmissions by these devices are not adequately controlled. Although V2V congestion control has been extensively investigated in the literature, such solutions may not be entirely applicable to V2P systems due to the unforeseeable and complex movement patterns of pedestrians. Furthermore, the level of pedestrian density in a given area could be higher by orders of magnitude in comparison with vehicles.

Power consumption, which primarily happens due to the high battery usage rate of GPS receivers and communication radios, is a sole issue of smartphones. Therefore, power consumption control mechanisms are required to reduce battery usage for V2P systems.

In our power consumption control approach, we first leverage the context awareness capability of smartphones to minimize the active duty cycle of the GPS receiver. Therefore, the GPS receiver is turned on or off based on the currently available context information. Whenever the device is in any of the following conditions the GPS should be turned off:

- Stationary (no Gyro variations)
- Indoor (inside building area or no GPS signal)
- Inside vehicle (moves as fast as the vehicle)
- Hiking trails, parks, and regions with no nearby vehicles

The change of these conditions is also detectable by smartphones. Moreover, the DSRC radio should also be turned off in these conditions, which also helps the congestion problem. Furthermore, the GPS duty cycle could be adjusted based on the required position update rate, which is controlled by our congestion control.

Our congestion control approach adaptively adjusts the transmission rate not only based on the explained context awareness scheme but also considering the surrounding conditions such as existence and distance of an approaching vehicle. Therefore, whenever the above-mentioned context conditions are not met, the smartphone only listens to the channel and receives BSMs from vehicles in order to warn the pedestrian in unpredicted situations. Furthermore, it should start the periodic transmission of its state information if an approaching vehicle is detected. Additionally, transmission range should also be controlled with the same criteria. For example, the VRU may reduce its transmission range if it senses many VRUs and few vehicles in its surroundings in order to form clusters and mitigate congestion. Since faster VRUs are more in risk of collision, they should transmit with a higher rate and power. For instance, a cyclist should transmit its position information further and more frequent in comparison with a pedestrian. However, exceptions should be made for specific users at higher risk such as public safety and road workers. An evaluation of the proposed congestion mitigation method using ns-3 is presented as a part of the following section.

V. OUTDOOR TEST AND SYSTEM EVALUATION

We designed, implemented and evaluated the proposed V2P framework with a DSRC-equipped Hyundai-Kia Sonata and a smartphone which had been DSRC-enabled by Qualcomm (Figure 1). Due to occupational safety considerations, in all of our tests, the smartphone was placed atop of a tripod with LOS of the vehicle to resemble a stationary pedestrian, except the proof-of-concept demonstration (Figure 5), which was performed with a real moving pedestrian and a slow speed vehicle (30 km/h). The tests were conducted in a small parking lot with a square size of approximately $500 \times 100 \ m^2$, on the West Virginia University campus. The position of a pedestrian in term of latitude and longitude for the stationary pedestrian is averaged over time. This position is then kept constant during the tests. We also benchmarked our averaged position with an accurate position information obtained from a differential GPS device and the error was not significant as we performed our tests in an open sky parking lot without GPS-blocking buildings.

Take into consideration that stationary pedestrian scenarios are more challenging as the GPS heading is not reliable. In non-stationary scenarios, the position and heading of the pedestrian can always be refined and filtered based on the path history points. However, the projection of the pedestrian path becomes difficult when the pedestrian is not moving, since the heading is a major contributor to the tracking module for pedestrian path prediction. In our tests no heading information is provided to the tracking module.

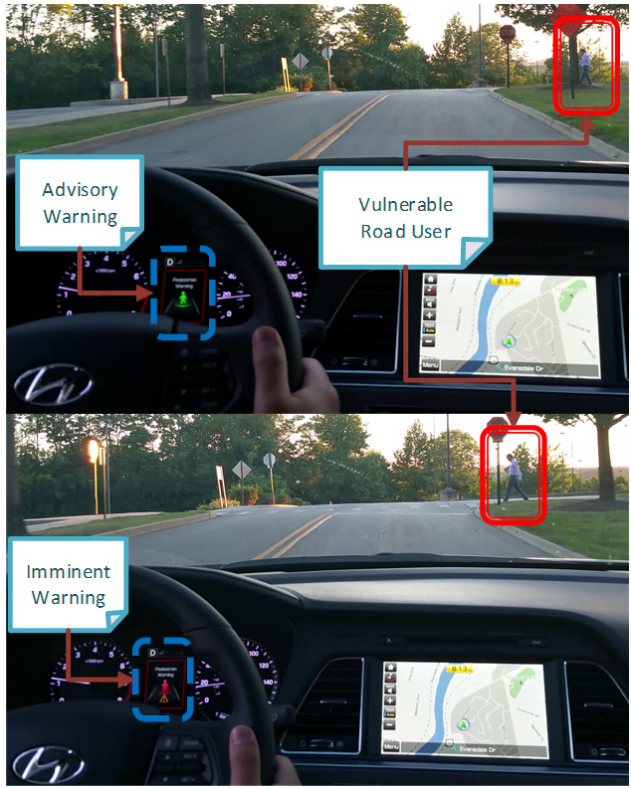

Figure 5- Proof-of-concept test of our cooperative VRU safety framework

The fastest scenario was around 70 km/h, which gave the safety system around 10 seconds of travel time at 200 meters of distance. This travel time was sufficient for generation of both advisory and imminent warnings (Figure 5). The imminent warning in our test has an audible beeping in addition to the cluster warning sign. Moreover, the pedestrian is only informed of imminent warning situations in all visual, auditory and haptic forms (Figure 1). Different vehicle speeds were tested to study path prediction and position estimation.

In these tests, the transmission power was set to 10 dBm and 20 dBm for the smartphone and the WSU, respectively. These power values give roughly 500 meters of two-way situational awareness range. The driver reaction time was set to $T_{DRD} = 2.5s$ and the maximum braking deceleration $d_{max} = -5.308 - 0.086 \times v_{brk}$ [33]. Moreover, the other two parameters of the application are set as follows $T_{guard} = 1s$ and $T_{mod} = 2s$.

Figure 6 illustrates the relative path of the smartphone from vehicle's real-time map perspective in a straight going vehicle scenario. Since the origin of the ENU coordinates of the vehicle are always calculated based on the center of the vehicle on the ground, even though the smartphone is stationary, the vehicle observes it as if it is approaching in the reverse direction. Different colors are used to draw the location of a smartphone, whenever it is classified inside one of the mentioned zones.

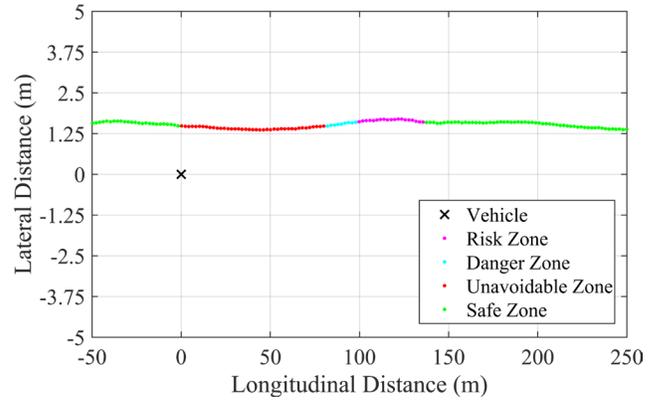

Figure 6- Relative path of the smartphone from the vehicle perspective.

The same scenario from the smartphone point-of-view is also shown in Figure 7. Similarly, whenever the smartphone detects itself inside one of the zones in front of the vehicle, it indicates the location of the vehicle with the color of that zone.

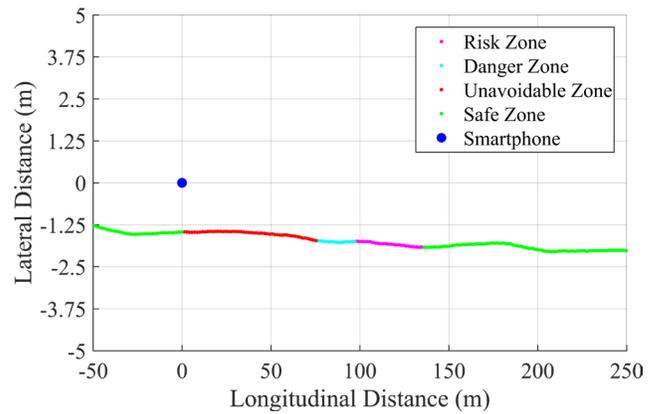

Figure 7- Relative path of the vehicle from the smartphone perspective.

The corresponding distance-to-stop of each zone along with the longitudinal distance of vehicle and smartphone for the whole scenario duration is depicted in Figure 8. The advisory and imminent warning generations are initiated whenever the $D_{lon} < DTS_{mod}$ and $D_{lon} < DTS_{guard}$ conditions are met for the first time, respectively.

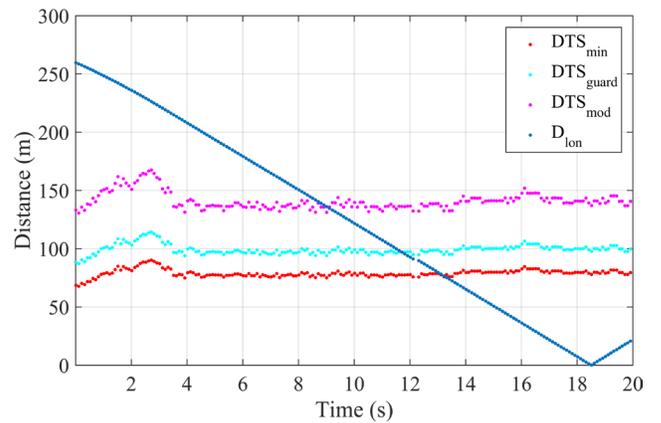

Figure 8- Corresponding distance-to-stop of each zone along with the longitudinal distance of vehicle and smartphone.

Equivalent figures for the curved road scenario are shown in Figure 9, Figure 10, and Figure 11.

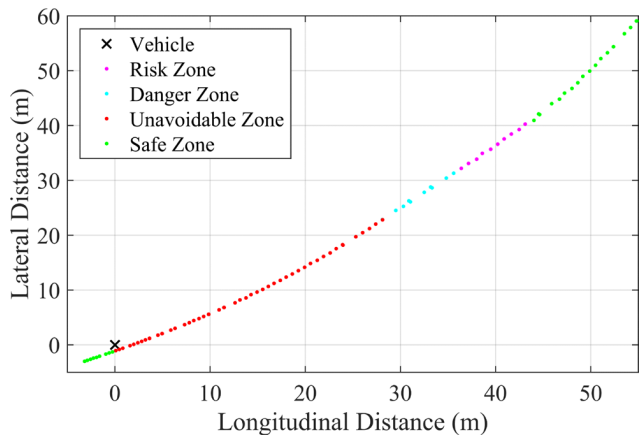

Figure 9- Relative path of the smartphone from the vehicle perspective.

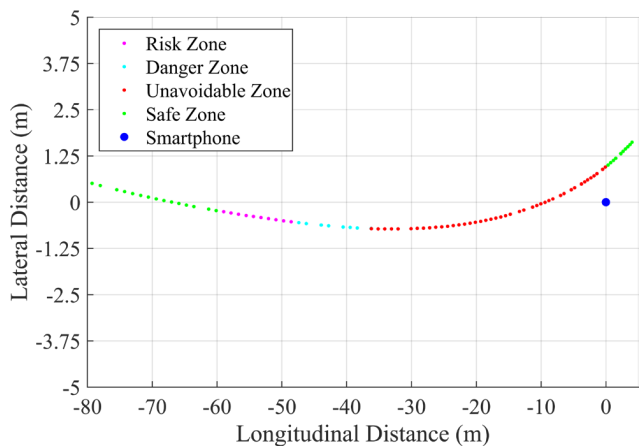

Figure 10- Relative path of the vehicle from the smartphone perspective.

Since distance-to-stop is a function of current speed and acceleration, its values are changing in a decreasing manner as the vehicle brakes at the end of curved road track in Figure 11.

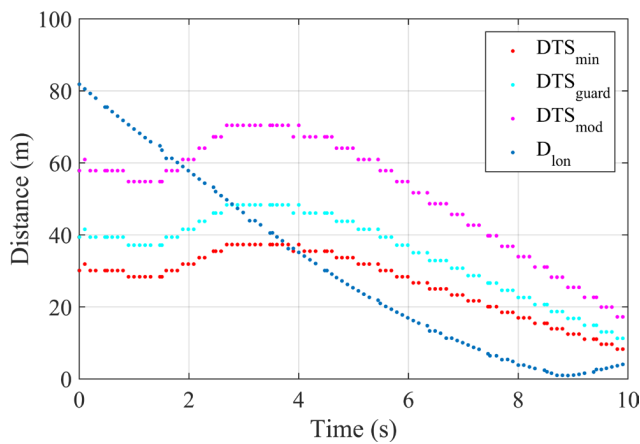

Figure 11- Corresponding distance-to-stop of each zone along with the longitudinal distance of vehicle and smartphone.

Furthermore, we use ns-3, which is a multipurpose full-stack network simulator, to evaluate the performance of our congestion mitigation mechanism. The ns-3 includes a complete implementation of DSRC standards [34]. In our simulations, the PHY layer is configured based on 802.11p protocol with 10MHz bandwidth and OFDM bit rate of 6Mbps. Other constant parameters are summarized in Table 2. In order to study the performance of the proposed context awareness mechanism, the transmission rate and range are fixed at 10 Hz and 500 meters, respectively. In the designed congestion scenario, 400 pedestrians are randomly distributed on both sides of a 1200 meters unsignalized straight road and moving across the sidewalks, while four vehicles drive from one end to the other.

Table 2- ns-3 Simulation Parameters

| Parameter | Value | Parameter | Value |
|---|---|---|---|
| No. Pedestrians | 400 | AIFSN | 7 |
| No. Vehicles | 4 | Contention Window | 15 |
| Transmission Rate | 10 Hz (Fixed) | Simulation Time | 50 sec |
| Range | 500 meters | Vehicle Speed | 24 m/s |
| Vehicle Tx Power | 20 dBm | Phone Tx Power | 10 dBm |

The scenarios with and without congestion control are compared regarding Packet Error Ratio (PER) in Figure 12. Note that the recorded PER versus distance is averaged for each device, smartphones, and vehicles, over the entire simulation time.

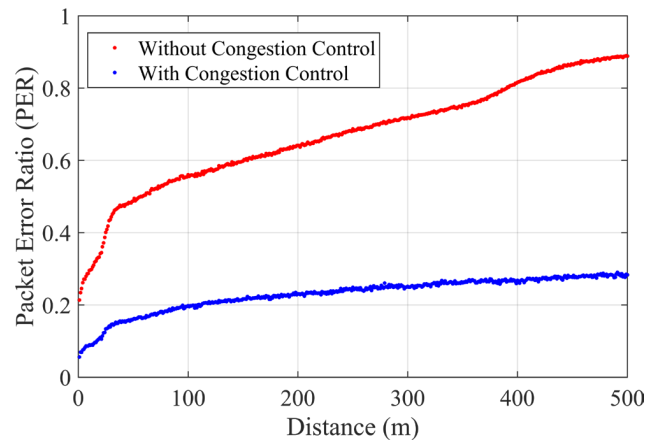

Figure 12- Congestion control performance in terms of PER versus distance.

Channel Busy Percentage (CBP) is also recorded in each device during the simulation time. The averaged CBP versus time for both scenarios is shown in Figure 13.

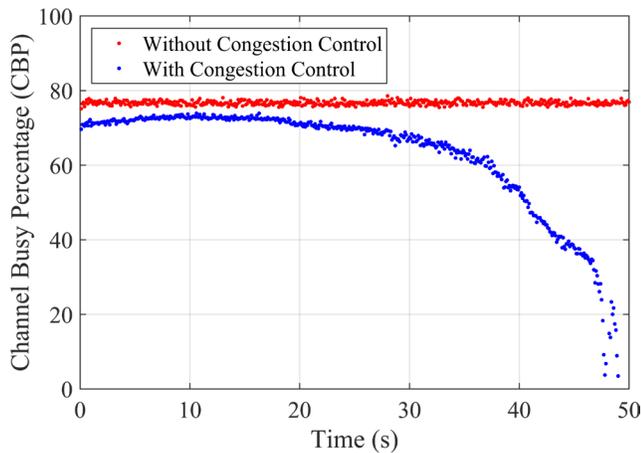

Figure 13- Averaged CBP versus time for with and without congestion control scenarios.

## VI. SUMMARY AND FUTURE WORKS

This article presented design, implementation, testing, and evaluation of our proposed vehicle to the pedestrian safety system. Particular characteristics of VRU movement dynamics, response times, and most fatal crash scenarios were considered in this framework to extend situational awareness and hazard detection capabilities. Furthermore, the proposed framework was designed to satisfy the scalability requirements of cooperative vehicle-VRU networks. Thus, power consumption and channel congestion mechanisms were described based on the particular needs of such networks. The reported practical tests and analyses indicated that our design is a promising VRU crash detection and mitigation solution. Our future work includes large-scale trials and empirical parameter optimization of proposed congestion control and power consumption control mechanisms.

Although the presented system design seems to provide a cornerstone for VRUs to talk to vehicles for safety, the realization of V2P safety systems is still facing many challenges. Further improvements and investigations in many aspects, such as network congestion, power consumption, localization, security, and availability of DSRC radio in a smartphone, are still required. Moreover, probabilistic approaches could be utilized to compensate for localization accuracy and human factors and their performance study is an interesting subject for further investigation.

ACKNOWLEDGMENT

The authors gratefully wish to acknowledge the Qualcomm Company for providing the DSRC-enabled smartphone, which has been utilized for the implementation and test phases of the proposed V2P safety system framework in this project.

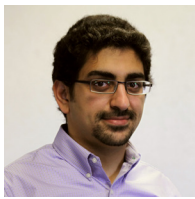
**Amin Tahmasbi-Sarvestani** received the B.S. degree in software engineering from the University of Isfahan, Isfahan, Iran, in 2009 and the M.S. degree in computer engineering (artificial intelligence) from Sharif University of Technology, Tehran, Iran, in 2011.

He is currently working toward the Ph.D. degree in computer science with the Lane Department of Computer Science and Electrical Engineering, West Virginia University, Morgantown, WV, USA. His research interests include connected and automated vehicles and vehicular networks.

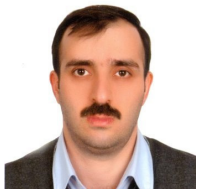
**Hossein Nourkhiz Mahjoub** is a Ph.D. student in electrical engineering with the Department of Electrical and Computer Engineering, University of Central Florida, Orlando, FL, USA. He received the B.S. and M.S. degrees in electrical engineering (Systems Communications) from University of Tehran, Tehran, Iran, in 2003 and 2008, respectively. He has more than nine years of work experience in the telecommunications industry before starting his Ph.D. in 2015. His research interests include wireless channel modeling, stochastic systems analysis, and vehicular ad-hoc networks.

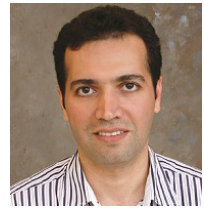
**Yaser P. Fallah** received his PhD from the University of British Columbia, Canada, in 2007. He is currently an associate professor of Electrical and Computer Engineering at University of Central Florida. From 2011 to 2016, he was an assistant professor at West Virginia University.

Prior to joining WVU, he worked as a research scientist at the University of California Berkeley, Institute of Transportation Studies (2008-2011). Dr. Fallah is an editor of IEEE Trans. On Vehicular Technology; he has served as the program chair of IEEE Wireless Vehicular Communication symposia in 2011 and 2014 and demo/poster co-chair of IEEE Vehicular Networking Conference in 2016. His current research, sponsored by industry and USDoT projects as well as NSF CAREER award, focuses on intelligent transportation systems and involves analysis and design of automated and networked vehicle safety systems.

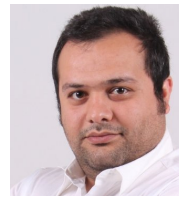
**Ehsan Moradi-Pari** received M.Sc. degrees in Electrical Engineering (Automatic Control) from the Sharif University of Technology in 2011 and the Ph.D. degree in electrical and computer engineering from West Virginia University, in 2016.

His research interests include modeling and analysis of Intelligent Transportation System, Vehicular Networks, Energy Cyber-Physical Systems, model predictive control, and cooperative control.

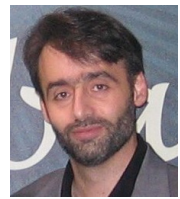
**Oubada Abuchaar** is a Senior Research Engineer working with the Advanced Research Team at Hyundai America Technical Center, Inc. (HATCI). He has been working with Hyundai since 2011. He participated in V2V-MD activities, and then led the Hyundai V2V-ICA applications development.

His current responsibilities are working on V2I research projects (V2I – Safety Applications, & CACC).

He brings 18 years of diverse Engineering experience in System Engineering, Product Development, Validation, and Quality. He worked with Ericsson Telephone Company on the installation of one million land phone lines in many cities across Syria. He developed and launched different radio head units for Daimler Chrysler vehicles for North America and Europe regions. He helped Ford to launch their first SIRIUS Satellite Radio Receiver when he worked with Delphi as a Resident System Engineer. He was among Clarion's Engineering team who developed and launched Ford's Navigation Head unit which was rated number one by JD Power for the year 2009.

Mr. Abuchaar holds BA in Electrical Engineering from Damascus University - Syria, and MS in the Science of Computer and Electrical Engineering from University of Detroit Mercy - MI.